# Who benefits from altmetrics? The effect of team gender composition on the link between online visibility and citation impact[1]

Orsolya Vásárhelyi[1] and Emőke-Ágnes Horvát[2]

[1] orsolya.vasarhelyi@uni-corvinus.hu
Laboratory for Networks, Technology and Innovation, Centre for Advanced Studies, Corvinus University, Budapest (Hungary)

[2] a-horvat@northwestern.edu
Department of Communication Studies, School of Communication, Northwestern University, Evanston, IL (USA)
Department of Computer Science, McCormick School of Engineering, Northwestern University, Evanston, IL (USA)
Northwestern University Institute on Complex Systems (NICO), Evanston, IL (USA)

**Abstract**
Online science dissemination has quickly become crucial in promoting scholars' work. Recent literature has demonstrated a lack of visibility for women's research, where women's articles receive fewer academic citations than men's. The informetric and scientometric community has briefly examined gender-based inequalities in online visibility. However, the link between online sharing of scientific work and citation impact for teams with different gender compositions remains understudied. Here we explore whether online visibility is helping women overcome the gender-based citation penalty. Our analyses cover the three broad research areas of Computer Science, Engineering, and Social Sciences, which have different gender representation, adoption of online science dissemination practices, and citation culture. We create a quasi-experimental setting by applying Coarsened Exact Matching, which enables us to isolate the effects of team gender composition and online visibility on the number of citations. We find that online visibility positively affects citations across research areas, while team gender composition interacts differently with visibility in these research areas. Our results provide essential insights into gendered citation patterns and online visibility, inviting informed discussions about decreasing the citation gap.

**Introduction**
Despite evidence that gender-diverse scientific teams produce less biased research (Campbell, Mehtani, Dozier, & Rinehart, 2013; Murray et al., 2019; Yang et al., 2022) and that women increase the overall intelligence of co-author teams (Bear & Wolley, 2011), female scholars earn less than their male colleagues (Ceci, Ginther, Kahn, & Williams, 2014; Shen, 2013), obtain less research funding (Ley & Hamilton, 2008; Van der Lee & Ellemers, 2015), receive less recognition (Caplar, Tacchella, & Birrer, 2017; Moss-Racusin, Dovidio, Brescoll, Graham, & Handelsman, 2012; Sarsons, 2017), and are less likely to be promoted (Régner, Thinus-Blanc, Netter, Schmader, & Huguet, 2019; Sarsons, Gërxhani, Reuben, & Schram, 2021; Weisshaar, 2017). Although recent years have seen proactive conversations about the underlying factors of female underrepresentation and the leaky pipeline in academia, progress toward equity for women in science has been slow (Sugimoto & Larivière, 2023).

---

[1] Please cite as: O. Vásárhelyi and E. A. Horvát. Who benefits from altmetrics? The effect of team gender composition on the link between online visibility and citation impact. Proceedings 19th International Society of Scientometrics and Informetrics Conference, Bloomington, Indiana, 2023.

Among the key issues that have been shown to impede female scientists' advancement at various career stages are: (1) a gender-based citation gap, meaning that male authors' articles receive more citations (Caplar, Tacchella, & Birrer, 2017; Dworkin et al., 2020; Ferber & Brün, 2011; Huang, Gates, Sinatra, & Barabási, 2020; Jadidi, Karimi, Lietz, & Wagner, 2018; Larivière, Ni, Gingras, Cronin, & Sugimoto, 2013; Lerchenmueller, Sorenson, & Jena, 2019; Maliniak, Powers, & Walter, 2013; Schrouff et al., 2019; Thiem, Sealey, Ferrer, Trott, & Kennison, 2018); (2) a bias in visibility (Leahey, 2007; Nittrouer et al., 2018) which also extends to the online dissemination of scholars' work (Procter et al., 2010; Shema, Bar-Ilan, & Thelwall, 2012; Sugimoto, Work, Larivière, & Haustein, 2017; Vásárhelyi, Zakhlebin, Milojević, & Horvát, 2021); and (3) an imbalance in credit-sharing between co-authors of research articles (Sarsons, 2017). Disparities in credit-sharing are driven by the gender-homophily embedded in men's citation practices (Dworkin et al., 2020; Ghiasi, Mongeon, Sugimoto, & Larivière, 2018) and the so-called "Matilda effect," which denotes a hesitance to acknowledge the achievements of female scientists and a tendency to credit male colleagues instead (Feeley, 2022; Patel, 2021).

Although citations and research visibility are essential for securing resources, awards, and promotion, scientists lack an understanding of their connection in the context of growing collaborative research efforts (Guimera, Uzzi, Spiro, & Amaral, 2005; Milojević, 2014). Science dissemination, which is increasingly happening online (Sugimoto, Work, Larivière, & Haustein, 2017), can help scholars reach a wider audience and raise awareness about their work among colleagues and relevant research communities. Sharing research findings on social media could aid underrepresented groups' dissemination efforts by avoiding potential gatekeepers, such as publishers and conference organizers (Yammine, Liu, Jarreau, & Coe, 2018). Research has shown a positive, albeit mostly weak, link between online visibility (e.g., attention to articles on social media) and citation impact (Bardus et al., 2020; Chang, Desai, & Gosain, 2019; Costas, Zahedi, & Wouters, 2015; Luc et al., 2021; Thelwall, Haustein, Larivière, & Sugimoto, 2013). However, it remains unexplored how the first and last authors' genders potentially influence the link between online visibility and citation impact. Addressing this gap empirically based on comprehensive data from the three broad research areas with distinct scholastic traditions (Computer Science, Engineering, and the Social Sciences) is especially timely now, as the research communities engage in productive discussions about the role of and best practices for public scholarship (Gilbert et al., 2020), and considers prominent issues of gender and marginalization (D'Ignazio, Graeff, Harrington, & Rosner, 2020; Fox, Menking, Steinhardt, Hoffmann, & Bardzell, 2017).

In this article, we compile a unique dataset from three different sources. First, we collect all research articles recorded in the Web of Science[2] that belong to the broad research areas of Computer Science, Engineering, or Social Sciences. We then gather how many times these articles were shared on any of the 14 major online platforms tracked by Altmetric[3]. These platforms include social media sites (e.g., Twitter, Facebook, Reddit), online news, knowledge repositories like Wikipedia, policy documents, and research blogs. Finally, we collect information from the Open Academic Graph about the number of citations these articles received within five years of their publication, as well as additional metrics, such as the size of these articles' co-author team and each author's h-index at the time of publication.

---

[2] https://clarivate.com/webofsciencegroup/solutions/web-of-science/

[3] https://www.altmetric.com/

To properly investigate the causal link between online visibility (quantified by the total number of article-shares online) and ensuing citation impact, we apply a statistical matching technique called Coarsened Exact Matching (CEM) (Iacus, King, & Porro, 2012). This method helps us identify statistical twins, i.e., pairs of similar articles where one was successful online and the other was not. Using the CEM method in this context is critical, since existing research on the relationship between online visibility and citation impact is correlational (Costas, Zahedi, & Wouters, 2015; Thelwall, Haustein, Larivière, & Sugimoto, 2013). Further controlling for several factors known to impact citations, we were able to isolate the direct effects of online visibility, team gender composition, and their interaction via regression models run on these matched data. Cognizant of the various issues with name-based algorithmic gender inference, we devised a new method to emulate errors in gender labeling and showed that our models are robust to the likely error level in this approach. We found support across the three studied research areas for the positive effect of online visibility on citation impact. Our models with various controls point to a citation benefit for teams with female first and/or last authors, as compared to articles with male first and last authors. We also find that this result depends on the research area, which invites discussions about custom strategies that best fit efforts to promote citation equity.

**Related work**
Although women's share in the US academic workforce is approaching 50%, women are still a minority among tenured faculty. In 2020, only 35% of the full professors were women (Education, 2020). Research has shown that women are less likely to pursue academic careers and are more likely to drop out of graduate schools (Ahuja, 2002; Kulis, Sicotte, & Collins, 2002). Even when they opt for an academic career path, female scientists earn less (Ceci, Ginther, Kahn, & Williams, 2014; Shen, 2013) and have access to less funding (Ley & Hamilton, 2008; Van der Lee & Ellemers, 2015). They are less likely to be promoted (Régner, Thinus-Blanc, Netter, Schmader, & Huguet, 2019; Sarsons, Gërxhani, Reuben, & Schram, 2021; Weisshaar, 2017), have fewer co-authors (Jadidi, Karimi, Lietz, & Wagner, 2018), are less likely to develop long-lasting scientific collaborations (Jadidi, Karimi, Lietz, & Wagner, 2018; Zeng et al., 2016), benefit less from co-authorship (Sarsons, 2017), publish less (Larivière, Ni, Gingras, Cronin, & Sugimoto, 2013; West, Jacquet, King, Correll, & Bergstrom, 2013), and publish in less prestigious venues (Holman, Stuart-Fox, & Hauser, 2018). Finally, their work receives fewer citations (Larivière, Ni, Gingras, Cronin, & Sugimoto, 2013).

Investigating factors associated with scientific success has become a popular research area over the last decade (Sinatra, & Barabási, 2020; Fortunato et al., 2018; Clauset, Larremore, & Sinatra, 2017; Sinatra, Wang, Deville, Song, & Barabási, 2016; Huang, Gates, Sarigöl, Pfitzner, Scholtes, Garas, & Schweitzer, 2014; Deville et al., 2014). Several studies have used large-scale publication data to examine the differences between male and female scientists' performance and achievements, measured by publications and citations, finding that publishing and citation practices are gendered (Arensbergen, Weijden, & Van den Besselaar, 2012; Jadidi, Karimi, Lietz, & Wagner, 2018; Zeng et al., 2016). Women are more likely to be associated with performing less prestigious tasks in clinical research (Macaluso, Larivière, Sugimoto, & Sugimoto, 2016), and articles with male first- and last-authors tend to receive more citations (Caplar, Tacchella, & Birrer, 2017; Dworkin et al., 2020; Ferber & Brün, 2011; Lerchenmueller, Sorenson, & Jena, 2019;

Maliniak, Powers, & Walter, 2013; Schrouff et al., 2019; Thiem, Sealey, Ferrer, Trott, & Kennison, 2018). Dworkin et al. found that female-led teams' under-citation is increasing over time, even as more women publish (Dworkin et al., 2020).

Science dissemination could contribute to mitigating the citation gap as higher visibility after publication might facilitate the accumulation of citations (Milojević, 2020a). Increasingly, science dissemination is happening online (Costas, Zahedi, & Wouters, 2015; Haustein, Peters, Sugimoto, Thelwall, & Larivière, 2014; Sugimoto, Work, Larivière, & Haustein, 2017; Zakhlebin & Horvát, 2020). Online public scholarship can be a rewarding experience for scientists that results in community engagement and vital scientific debate, and it can increase awareness about one's work (Priem & Hemminger, 2010; Yammine, Liu, Jarreau, & Coe, 2018). The correlation between online visibility and eventual citation impact has been studied via correlational and experimental means (Bardus et al., 2020; Costas, Zahedi, & Wouters, 2015; Luc et al., 2021; Thelwall, Haustein, Larivière, & Sugimoto, 2013). These studies found a weak positive link between online visibility and subsequent citations.

The online visibility of scientists is not gender-neutral (Paul-Hus, Sugimoto, Haustein, & Larivière, 2015; Vásárhelyi, Zakhlebin, Milojević, & Horvát, 2021). For instance, it has been shown that scientific communication on social media is disproportionately male-dominated (Procter et al., 2010; Sugimoto, Work, Larivière, & Haustein, 2017). Self-promotion is crucial online, but women typically avoid it due to backlash against non-gender conforming confidence and assertiveness (Moss-Racusin & Rudman, 2010; Peng, Teplitskiy, Romero, and Horvát, 2022). Men also blog more (Shema, Bar-Ilan, & Thelwall, 2012), are more likely to serve as gatekeepers on social media (Matias, Szalavitz, & Zuckerman, 2017), and edit Wikipedia at a higher rate (Collier & Bear, 2012; Forte et al., 2012; Hargittai & Shaw, 2015; Hill & Shaw, 2013; Lam et al., 2011). These differences may perpetuate and potentially widen gender gaps since, for instance, the underrepresentation of women among Wikipedia contributors is often reflected in biased content against women (Reagle & Rhue, 2011).

Recently, the Economics (Hengel & Moon, 2020; Koffi, 2021), Neuroscience (Dworkin et al., 2020), and Astrophysics (Caplar, Tacchella, & Birrer, 2017) fields have explored the citation disadvantage of female-authored articles. We lack similar investigations in Computer Science, Engineering, and Social Sciences regarding female scholars' representation and the adoption of online science dissemination among these scholars. Computer Science, Engineering, and Social Sciences are different fields, which provides us with a broad basis for highlighting various patterns in scientific production, dissemination, and impact. Focusing on these three domains, our article uncovers how team gender composition interacts with research visibility in determining citation impact.

**Data and methods**

*Data: Combining Altmetric.com, the Web of Science, and the Open Academic Graph*
This study's empirical basis is a database containing scientific articles published in 2012 and recorded on Altmetric.com (Altmetrics, n.d.) as having been mentioned online on one of the 14 major platforms tracked by the service. Altmetric.com registers mentions of scientific articles in public social media posts (e.g., Twitter, Facebook, Reddit), online news, Wikipedia, policy documents, and research blogs. There are 1,101,076 scholars whose articles were mentioned on

Altmetric in 2012. Based on unique DOIs (document object identifiers), we connect information on these articles and their authors with a dataset that consists of all articles published in 2012, according to the Web of Science (WoS); 1,823,069 articles in total. To ensure that we capture meaningful co-author team dynamics, we hereafter consider only articles with fewer than 10 authors, which leaves us with 241,386 articles written by 537,486 scholars from a wide range of research domains. To identify broad research areas and subfields, such as Anthropology, Robotics, or Information Systems, we apply a classification method that uses each article's references to infer the topic of a bibliographic item (Milojević, 2020b). This method considers Web of Science subject categories as units of classification but relies only on references published in journals with a single subject category (i.e., not "multidisciplinary"). The approach classifies papers into exclusive subject categories and an aggregated broad research area. We then obtained the number of citations these articles received by the end of 2017 from the Open Academic Graph. This article focuses on three broad research areas: Computer Science, Engineering, and Social Sciences.

*Gender inference and its evaluation*
We assign authors' gender based on their first names by applying the "simple gender-guesser" inference algorithm developed by Ford et al. (Ford, Harkins, & Parnin, 2017), which is an updated version of "gender-computer" developed by Vasilescu et al. (Vasilescu, Capiluppi, & Serebrenik, 2014). These algorithms use the frequency of male and female first names from national statistics in 68 different countries. They have a conservative approach, leaving ambiguous cases as non-identifiable. The inferred "gender" does not refer to the scholars' self-chosen gender. Instead, it maps onto the gender expectations created by the name, which might trigger (un)conscious biases that could affect online visibility and citations. An important limitation of this gender assignment method is that it is not suited for non-binary, transgender, and intersex identities. We discuss the shortcomings of our approach in Section 5.2. Ford et al.'s procedure results in 13,678 men and 4,655 women, and 34% unknowns in Engineering, 28% unknowns in Computer Science, and 27% unknowns in the Social Sciences, respectively. We filter out articles for which the first or last authors' gender cannot be inferred. We are left with 20,186 articles from the Web of Science that had also been recorded by Altmetric as having at least one mention online in 2012 (see Table 1).

We group articles by the gender combination of the articles' first and last authors, resulting in four distinct categories. FF denotes articles with female first and last author; FM denotes female first and male last author; MF denotes male first and female last author; MM denotes male first and last author. Table 1 shows the gender composition of articles published in 2012 in the WoS and the sample shared online and registered on Altmetric.com. The Web of Science and Altmetric samples yield very similar gender team compositions. The difference between the two data sources never exceeds 5%; therefore, Altmetric is a subsample comparable to Web of Science. As expected, based on the female representation of the three analyzed broad research areas, the Social Sciences have the highest percentage of FF teams (Web of Science 25.5% and Altmetric 24.1%). Engineering and Computer Science have similar percentages of FF teams on Web of Science (8.3% and 8.1%). The percentage of FF teams in Computer Science on Altmetric is higher than expected based on Web of Science (11.9% vs 8.1%), which indicates that a higher percentage of female Computer Scientists' work receives visibility online than the percentage of scholars who are publishing in this area. More than half of the articles in all research areas are published by MM teams. This percentage is the highest in Engineering (Web of Science 62.3% and Altmetric 59.1%) and the lowest in the Social Sciences (Web of Science 56.4% and Altmetric 54.3%). In

Engineering, teams with a female first author (FF or FM) have a higher percentage of representation on Altmetric than in the Web of Science.

Table 1. Descriptive statistics of our pre-processed data. Number of articles published in 2012 according to the Web of Science (WoS) and the number of articles with online shares as registered by Altmetric. Values are shown for three broad research areas. Articles are categorized based on the team's gender composition. FF denotes female first and last author, FM is female first and male last author, MF is male first and female last author, MM is male first and last author.

|  | *Computer Science* | | *Engineering* | | *Social Sciences* | |
| ---: | --- | --- | --- | --- | --- | --- |
|  | WoS | Altmetric | WoS | Altmetric | WoS | Altmetric |
| Number of articles | 24,254 | 2,689 | 174,273 | 6,685 | 42,278 | 10,812 |
| Articles with less than 10 authors | 14,328 | 1,368 | 75,166 | 5,463 | 36,281 | 7,633 |
| FF teams (%) | 8.1% | 11.9% | 8.3% | 8.5% | 25.5% | 24.1% |
| FM teams (%) | 14.4% | 14.3% | 18.3% | 20.7% | 10.0% | 11.7% |
| MF teams (%) | 11.2% | 12.7% | 11.2% | 11.7% | 8.1% | 9.9% |
| MM teams (%) | 66.2% | 61.1% | 62.3% | 59.1% | 56.4% | 54.3% |

Name-based gender-inferring approaches are less accurate on non-Western names (Karimi, Wagner, Lemmerich, Jadidi, & Strohmaier, 2016; Lockhart, King & Munsch, 2022). To account for potential mislabeling by the inference method, we quantify the level of error by comparing the output of the algorithm with manually assigned genders for 400 randomly selected articles, ensuring that we have 100 articles for each team gender composition (FF, FM, MF, MM; no articles with undecided authors were selected for the test). We look up the first and last authors for each sampled article and identify their gender based on the name and pictures found online. If the first or last author's gender was not unambiguously identifiable during the manual inspection, we classified the article as undecided. We also compared the algorithm that we used with the common gender inferring tool called "gender-guesser" in Python. Figure 1 shows the accuracy of the algorithm (Ford, Harkins, & Parnin, 2017) and "gender-guesser" compared to the manually labeled data.

The algorithm performs better than the "gender-guesser" in terms of both recall and F1-score. Yet, it still miscategorizes 3% of women in the Social Sciences (compared to 11.5% for "gender-guesser"), 9.75% in Computer Science ("gender guesser" error is 26.5%), and 13.75% in Engineering ("gender guesser" error is 28.5%). This trend aligns with the expectation that the accuracy is higher across methods in broad research areas where Western names are more common.

To account for the error level of gender inference throughout our analyses, we artificially introduce random errors in the original gender inferences obtained using Ford et al.'s method. The magnitude of the introduced errors follows our error quantification (Figure 1). This step is essential to show that the results we report later are robust to the expected errors. We create randomized gender-swapped datasets where the percentages of the four gender compositions follow the manually inferred ones. Table 2 shows the team gender composition percentages based on the original data as per Ford et al.'s algorithm and the manually labeled genders. We count for each team's gender composition how many times the algorithm mislabeled it (for instance, marking it as FF instead of FM). Then, we swap team gender composition labels to account for the mismatches. For example, if 3 of 100 FF teams are wrongly categorized as FF instead of FM, we randomly replace the team labels for 3% of the FF teams with FM. We repeat this process for each category and record the ones we can not unambiguously decide as undecided. We follow this process 100 times for the three broad research areas. We run our models and statistics on the original dataset via Ford et al.'s algorithm and the randomized samples to ensure that the results are robust to the error rate of the algorithmically inferred genders.

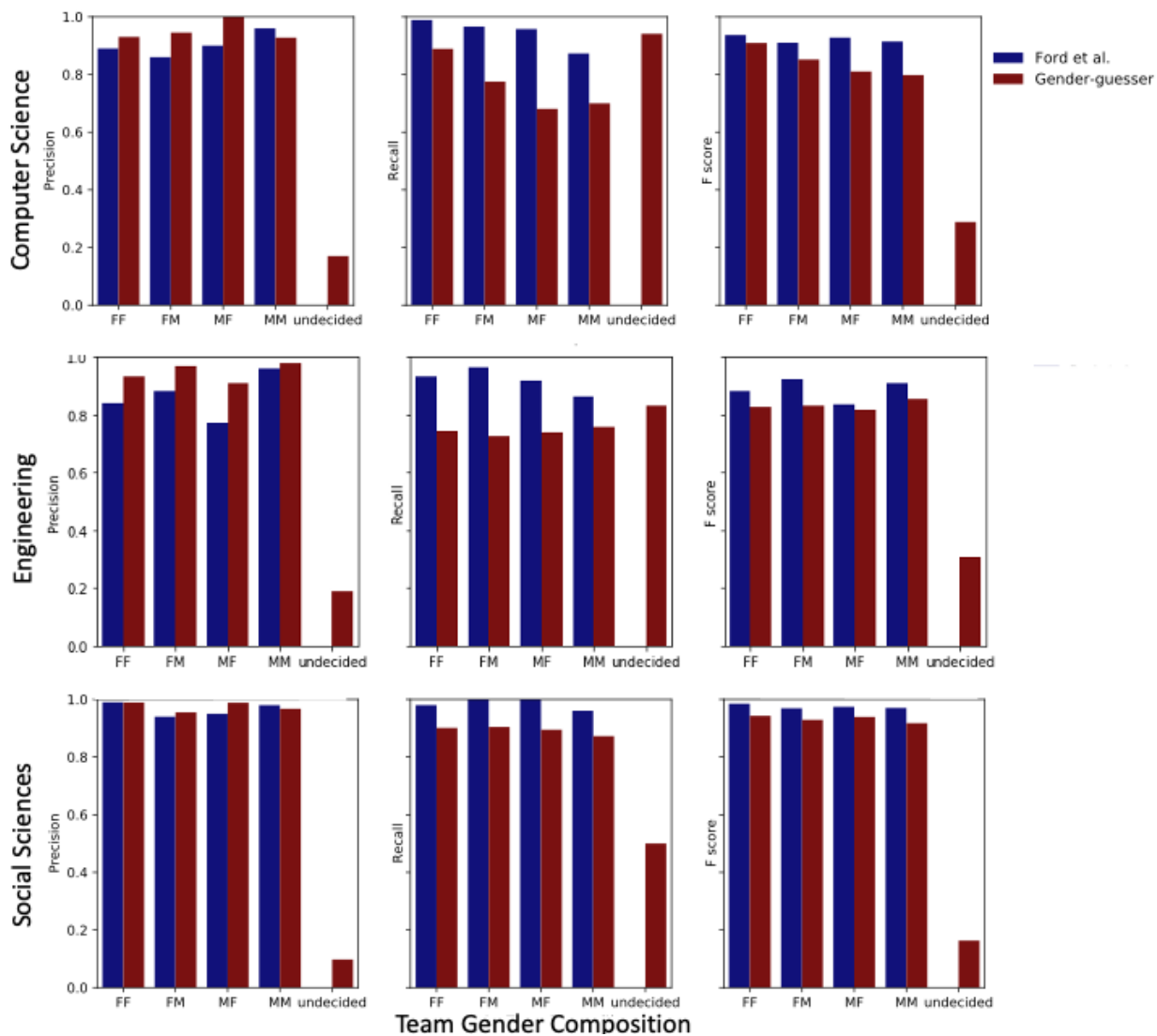

**Figure 1. Evaluating the gender inferring method by Ford et al. (Ford, Harkins, & Parnin, 2017) in comparison with the commonly used "gender-guesser." We compute the inference accuracy metrics**

(Precision, Recall, F-score) for each of the four gender compositions separately. Manually labeled data for 400 randomly selected articles served as the baseline.

Table 2. Percentage of articles with the given team gender compositions based on the Ford et al. algorithm using the original dataset and based on the manually assigned gender labels (gender-swapped samples). Gender-swapped datasets used for robustness checks follow the team gender compositions found manually.

|  | *Computer Science* | | *Engineering* | | *Social Sciences* | |
| --- | --- | --- | --- | --- | --- | --- |
|  | Original | Swapped | Original | Swapped | Original | Swapped |
| FF | 11.92% | 10.75% | 8.51% | 7.85% | 24.15% | 24.13% |
| FM | 14.25% | 12.65% | 20.74% | 18.49% | 11.70% | 10.99% |
| MF | 12.65% | 13.38% | 11.69% | 9.74% | 9.90% | 9.41% |
| MM | 61.18% | 60.67% | 59.06% | 58,97% | 54.25% | 53.74% |
| Undecided |  | 2.56% |  | 4.95% |  | 1.73% |

**Quantifying online visibility and citation impact**

To study the impact of different team compositions, we compute online visibility as the number of article shares recorded in Altmetric in the year the article was published and citations obtained within five years of publication. Here we consider the publication year 2012 and count the citations received by the end of 2017. Table 3 shows the 90$^{th}$ and 95$^{th}$ percentiles and the maximum online visibility and citation impact for each broad research area. Visibility and citations have highly skewed distributions, much like other measures of success (Jadidi, Karimi, Lietz, & Wagner, 2018; May, Wachs, & Hannák, 2019; Vedres & Vasarhelyi, 2019). To define successful articles, we use hereafter the top 10% of the articles with the highest number of shares and the top 10% of the articles with the highest number of citations.

Table 3. 90th and 95th percentiles of online visibility and citation impact for each of the three broad research areas. We consider two types of success: online and based on citations. If the article is in the 90th percentile based on shares, we consider it to be successful online. Similarly, success based on citations means that the article is in the 90th percentile. Since the number of article shares within a year of publication have a skewed distribution, the 90$^{th}$ and 95$^{th}$ percentiles are low.

|  | *Online visibility (2012)* | | | *Citation impact (2017)* | | |
| --- | --- | --- | --- | --- | --- | --- |
|  | Top 10% | Top 5% | Maximum | Top 10% | Top 5% | Maximum |
| *Computer Science* | 4 | 8 | 372 | 98 | 170 | 728 |
| *Engineering* | 2 | 6 | 696 | 69 | 124 | 1,032 |
| *Social Sciences* | 4 | 10 | 1,832 | 198 | 296 | 5,808 |

*CEM: Controlling for aspects of scientific production that might influence citations*

To accurately assess the link between online visibility and citations for different team gender compositions, we need to control for three crucial factors that the literature has identified as influencing citations. First, scientific success has been shown to follow a cumulative advantage process where previously successful scholars garner an increasing fraction of citations (Milojević, 2020a). To take this rich-get-richer effect into account, we track the *highest h-index* in the co-author team, since h-index is the most widely used measure of scientists' success (Bornmann & Daniel, 2005; Lehmann, Jackson, & Lautrup, 2006). Second, citations are influenced by the prestige of the publication venue, which is frequently quantified by the *impact factor* of the journal where the article was published (Sarigöl, Pfitzner, Scholtes, Garas, & Schweitzer, 2014; Sekara et al., 2018). Although h-index and journal impact factor have been heavily criticized for not being the most accurate and fair metrics (Bi, 2022; Teixeira da Silva, 2018), the scientific community still uses these metrics as a proxy to assess the influence and importance of an article or a journal (Roldan-Valadez, 2019), and to evaluate career advancement (Wang, 2022). Third, there is extensive work on the importance of social capital in academia (Arnaboldi, Dunbar, Passarella, & Conti, 2016; Katz & Hicks, 1997), which prompts us to control for the *size of the co-author team*, as more team members ought to promote the work more broadly.

We control for the potential influence of these three aspects on citations in a quasi-experimental setting using Coarsened Exact Matching (CEM) (Iacus, King, & Porro, 2012). This technique allows us to assess the direct effect of online visibility and team gender composition on citation impact by considering balanced data in terms of maximum h-index, journal impact factor, and co-author team size. The method involves a pre-processing of the data that creates a matched dataset based on these three aspects in order to estimate the so-called sample average treatment effect on the treated or "SATT" by decreasing the noise in the data (Iacus, King, & Porro, 2012; King, Nielsen, Coberley, Pope, & Wells, 2011). Our treatment variable is online visibility (i.e., belonging to the top 10% most shared articles in a broad research area in 2012). This means that the matched data has similar distributions in terms of highest h-index, impact factor, and co-author team size across the two groups (top 10% most shared articles and bottom 90%). The outcome variable is citation impact, which we quantify as the logarithm of the total number of citations plus one received within five years of article publication.

We run three different OLS regression models on the resulting matched database to quantify 1) the effect of online visibility on citations, 2) the effect of different gender team compositions on citations (compared to MM teams) while controlling for online success, and 3) the additional effects of interactions between online visibility and team gender composition on citations.

**Results**
When exploring the link between online visibility and citation impact without any controls or consideration of gender team composition, we find that there is, on average, a 24.38 citation difference (1.52 in log(citations+1)) between Computer Science articles that have high online visibility and those that do not. Similarly, the expected difference in the number of citations is 9.54 in Engineering (0.24 in log(citations+1)) and 48.19 in Social Sciences (1.52 in log(citations+1)). This indicates that articles in Social Sciences benefit most in terms of citations from being highly shared online.

To display whether gender team composition correlates with online visibility and citations, Figure 2 shows the number of successful teams by gender composition and success category. Error bars indicate the minimum and maximum times the different team compositions are in the top 10% based on the 100 gender-swapped samples. We test whether there are statistically significant differences between the success rates of the four team compositions (FF, FM, MF, and MM). Chi2 tests do not show significant relationships between articles' team gender composition and belonging to the top 10% of the most highly shared articles (according to Altmetric) in either of the three broad research areas. However, there are significant differences based on team gender composition in terms of being in the top 10% of the most highly cited articles in Computer Science (Chi2 test, $p_{CS}$=0.006) and Engineering (Chi2 test, $p_{Eng}$=0.000).

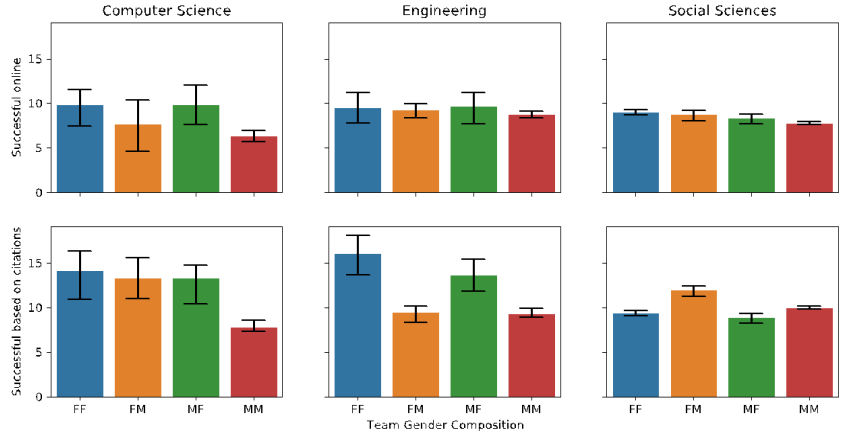

**Figure 2. Number of successful teams by gender composition in terms of online visibility (top) and citation impact (bottom). To quantify errors stemming from potential gender mislabeling, bars indicate the range of success rates based on 100 gender-swapped samples.**

To control for potential confounding factors, we perform CEM and match articles that have high visibility online to those that do not have high visibility but do have the same coarsened variable values in terms of maximum h-index, journal impact factor, and team size. We find matches for 85 out of 94 high-visibility Computer Science articles, 431 out of 444 high-visibility Engineering articles, and 612 out of 628 high-visibility Social Science articles. Table 4 shows the goodness-of-fit of the matched data by variables and overall. Overall, Social Sciences is matched best (L1=0.488) and Computer Science worst (L1=0.724), which is due to the higher imbalance in journal impact factor (0.259) and smaller sample size.

**Table 4. Univariate and Multivariate Imbalance metrics of Matched Data. The L1 imbalance measure looks at the entire joint distribution of the covariate space. L1=0 means fully balanced matching, and L1=1 means not balanced at all.**

|  | Computer Science | Engineering | Social Sciences |
|---|---|---|---|
| Maximum h-index | 0.155 | 0.077 | 0.080 |
| Journal impact factor | 0.259 | 0.056 | 0.058 |
| Team size | 0.077 | 0.042 | 0.036 |
| L1 | 0.724 | 0.491 | 0.488 |

On this matched data, we run three OLS regression models where the outcome variable is the logarithm of the number of citations in 2017. In the first model, we are interested in the direct impact of online visibility on citations, net of the three control variables (highest h-index in the co-author team, impact factor of the journal, and size of the co-author team). We add team gender compositions in the second model and estimate their effect on citations after controlling for online visibility. We also consider the interaction between online visibility and team gender composition in the third model. We run these models on the original dataset and the gender-swapped samples. Table 4 shows the results of the three models. Next we report the percentage of significant coefficients in each case.

The baseline model (Model 1) shows that online attention in the year when the paper was published (2012) significantly affects citations five years later (2017) in all of the three broad research areas. In the controlled setup of CEM, the coefficients are smaller than we found initially (i.e., 1.52 in Computer Science, 0.24 in Engineering, and 1.52 in Social Sciences). In Computer Science, articles that belong to the top 10% of most highly shared articles online receive 0.404 more log number of citations. In Engineering, this value is 0.216, and in Social Sciences, it is 1.309. These results hold across gender-swapped datasets.

With Models 2 and 3, we estimate the effect of team gender composition on citations while controlling for online visibility. Our baseline team gender composition is MM, i.e., teams with male first and last authors. Therefore, all coefficients indicate the difference in the log number of citations compared to MM teams.

In Computer Science, the effect of online visibility loses its significance in Model 2 and regains it if we add the interaction between online visibility and team gender composition (Model 3). This indicates that the impact of online visibility is not independent of team gender composition. In both Models 2 and 3, all team compositions have significant positive coefficients compared to MM, suggesting that articles produced by teams with female first and/or last authors receive more citations when all else is equal in terms of maximum h-index in the team, journal impact factor, team size, and online visibility. However, interaction terms in Model 3 have negative coefficients and are significant in the case of FF and MF teams. This indicates that Computer Science teams with female last authors benefit less from high visibility in terms of citations than MM teams, meaning that teams with female last authors obtain fewer citations than MM teams with the same visibility. The significance levels of non-interaction terms on citations are robust across gender-swapped teams in all cases. The negative impact of online visibility compared to MM teams was significant in 65% of the gender-swapped FF teams and in 13% of the MF teams.

In Engineering, the positive effect of online visibility on citations is significant, but only articles published by FF teams have a significant citation advantage compared to MM teams (Models 2 and 3). Robustness checks on Model 3 show that 63% of the gender-swapped datasets preserve the advantage of online visibility and 72% of the gender-swapped datasets preserve the advantage of FF teams. This indicates that Engineering teams' gender composition was only beneficial compared to MM teams for FF teams, who might benefit from online visibility in 19% of the cases. The positive effect of visibility on citations is also significant in Social Sciences, even after introducing team gender composition. Additionally, across Models 2 and 3, gender-diverse teams (FM and MF) are associated with more citations than MM teams. The interactions between team gender composition and online visibility are not significant. These results are robust to gender-swapping. In all cases, variables have the same significance levels in the gender-swapped samples as in the original data, which is explained by the high performance of the gender inference algorithm in the Social Sciences.

**Table 5. Linear regression models on data matched via CEM. Outcome variables are the logarithm of the number of citations in 2017 in the respective broad research areas. Coefficients indicate the sample average treatment effect on the treated (SATT). For Model 3, we show the percentage of gender-swapped datasets in which the corresponding coefficients were significant. Significance codes: '***' denotes p<0.001; '**' denotes p<0.01; '*' denotes p<0.05.**

| | Computer Science (N=585) | | | | | | |
|---|---|---|---|---|---|---|---|
| | Model 1 | | Model 2 | | Model 3 | | |
| | *Coef* | *p* | *Coef* | *p* | *Coef* | *p* | *% sign.* |
| Successful online (top 10% based on article shares) | 0.404 | * | 0.325 | | 0.737 | ** | 100% |
| FF | | | 0.812 | *** | 0.958 | *** | 100% |
| FM | | | 1.061 | *** | 0.958 | *** | 100% |
| MF | | | 0.720 | *** | 0.958 | *** | 100% |
| Top 10% * FF | | | | | -0.993 | *** | 65% |
| Top 10% * FM | | | | | -0.614 | | 0% |
| Top 10% * MF | | | | | -0.882 | * | 13% |
| (Intercept) | 2.225 | *** | 2.269 | *** | 2.217 | *** | 100% |

| | Engineering (N=4,273) | | | | | | |
|---|---|---|---|---|---|---|---|
| | Model 1 | | Model 2 | | Model 3 | | |
| | *Coef* | *p* | *Coef* | *p* | *Coef* | *p* | *% sign.* |
| Successful online (top 10% based on article shares) | 0.216 | ** | 0.213 | ** | 0.190 | * | 63% |
| FF | | | 0.227 | ** | 0.181 | * | 72% |
| FM | | | 0.052 | | 0.054 | | 0% |
| MF | | | 0.116 | | 0.124 | | 25% |
| Top 10% * FF | | | | | 0.435 | * | 19% |
| Top 10% * FM | | | | | -0.024 | | 0% |
| Top 10% * MF | | | | | -0.074 | | 0% |
| (Intercept) | | | 2.629 | *** | 2.588 | *** | 100% |

| | Social Sciences (*N*=6,919) | | | | | | |
|---|---|---|---|---|---|---|---|
| | Model 1 | | Model 2 | | Model 3 | | |
| | *Coef.* | *p* | *Coef.* | *p* | *Coef.* | *p* | *% sign.* |
| Successful online (top 10% based on article shares) | 1.309 | *** | 1.304 | *** | 1.244 | *** | 100% |
| FF | | | 0.162 | | 0.124 | | 0% |
| FM | | | 1.535 | *** | 1.558 | *** | 100% |
| MF | | | 1.332 | *** | 1.344 | *** | 100% |
| Top 10% * FF | | | | | 0.393 | | 0% |
| Top 10% * FM | | | | | -0.242 | | 0% |
| Top 10% * MF | | | | | -0.148 | | 0% |
| (Intercept) | 1.686 | *** | 1.328 | *** | 1.333 | *** | 100% |

## Discussion

*Summary and main findings*

In this article, we investigate how online sharing of scientific articles and team gender composition is linked with the number of citations articles receive five years after publication. Specifically, we collect data about *all* the scientific articles published and shared online in 2012 in three broad research areas (Computer Science, Engineering, and the Social Sciences). From Altmetric.com, we record how many times these articles are shared online in their year of publication and then connect them via DOIs to the Web of Science and the Open Academic Graph to collect additional metrics, such as the number of citations they received by 2017.

We infer the gender of the articles' first and last authors and categorize author teams into four distinct groups: female-female (FF), female-male (FM), male-female (MF) and male-male (MM). We find that 56.4%–66.2% of articles are published by MM teams. Teams with a female lead and/or last authors are generally better represented among the articles shared online than expected based on their publication rate. To account for potential errors in name-based gender inference, we first evaluate the algorithm's accuracy on a manually labeled sample. Then, we generate randomized gender-swapped datasets where the frequency of the four gender compositions follows the result of the manual labeling. We use these gender-swapped samples throughout our analyses to quantify the sensitivity of our results to the expected level of error in gender inference.

We apply Coarsened Exact Matching (CEM) to create a quasi-experimental setting. We run regression models to evaluate for each broad research area the effect of online visibility (defined as whether the article is in the top 10% of the most highly shared articles in 2012) and team gender composition on citation impact (top 10% most highly cited articles in 2017). This method allows us to adequately control for essential aspects of scientific success, such as the maximum h-index in the team, journal impact factor, and co-author team size, and to explore the relationship between online visibility and citation impact. We find that in all three research areas, the sample average treatment effect (SATT) of online visibility is a significant positive predictor of the log number of citations five years after publication. Using the CEM approach in this context is novel and essential, since most prior work on the relationship between online visibility and citation impact is based on correlational analyses (Costas, Zahedi, & Wouters, 2015; Thelwall, Haustein, Larivière, & Sugimoto, 2013).

The effect of field gender composition is research area dependent. We find robust evidence across gender-swapped datasets in Computer Science that teams with female first and/or last authors have higher online visibility than MM teams but that teams with female last authors benefit less from this higher visibility than MM teams. In Engineering, FF teams benefit more from online visibility compared to MM teams. However, the interaction is positive and significant in merely 19% of the cases. In the Social Sciences, team gender composition and online visibility are significant predictors of the number of citations. FM and MF teams have more predicted citations than MM teams. Yet, FF teams' predicted citation impact is not significantly different from MM teams' citation impact. This holds across all gender-swapped samples and indicates that team gender diversity could be an essential asset in the Social Sciences.

*Limitations and future work*
This work relies on data that do not contain individuals' self-reported gender. Even though we quantify and account for the errors introduced by algorithmic gender inference to maximize the reliability of our findings, several issues remain with the complicated problem of automated gender assignment. First, name-based gender inferring methods are not trans-inclusive, operating under the crude assumption of binary genders (Keyes, 2018). Second, these algorithms are also biased towards Western names (Karimi, Wagner, Lemmerich, Jadidi, & Strohmaier, 2016). Therefore, our findings cannot be generalized to non-Western authors. The relatively low rate of errors in the Social Sciences, as opposed to Computer Science and Engineering, indicates the presence of such bias.

Since we focus on understanding the citation impact of online visibility based on articles published each year, we can not analyze whether this link has changed over time. Furthermore, in recent years the conversation and movement to support underrepresented groups in academia have developed significantly (Patel, 2016; Perkel, 2020), which could have mitigated, for example, FF teams' disadvantage in comparison to MM teams in Computer Science. Additionally, social media has become a more integral part of scientists' promotion efforts (Sugimoto, Work, Larivière, & Haustein, 2017; Zakhlebin & Horvát, 2020). The higher emphasis on disseminating publications online might have implications that require further research.

Furthermore, we are aware that many other factors can impact how a scientific work is picked up by the scientific community, such as topic selection, authors' personal network, career mobility affiliation, and country of origin (Fortunato et al., 2018) just to mention a few that we can not control for in this study. Finally, this study does not aim to predict the number of citations a paper receives within a given time frame of its publication. Therefore, it does not consider several other factors relevant for citation impact, such as scientists' position in the full co-authorship network (Sarigöl, Pfitzner, Scholtes, Garas, & Schweitzer, 2014; Sekara et al., 2018) or the subfield and topic of the article (Tahamtan, Safipour Afshar, & Ahamdzadeh, 2016). Future work could build models that augment the ones presented here with a suite of structural and content-based factors.

**Conclusions**

Although in all three broad research areas examined here male-led teams are the majority, the difference in the representation of female-female teams is significant (12% in Computer Science, 9% in Engineering, and 24% in Social Sciences). Gender diversity could be beneficial (Campbell, Mehtani, Dozier, & Rinehart, 2013; Rock & Grant, 2016) in an inclusive environment (Ahuja, 2002; Joshi, 2014) where women and members of other underrepresented groups have an equal say.

Our main finding is that teams with high online visibility receive significantly more citations. However, team gender composition interacts differently with visibility in the three broad research areas studied here. Online visibility has the potential to mitigate the gender citation gap. Yet, in Computer Science, we find that teams with female last authors benefit less from higher online visibility than teams with male last authors. Prior studies about the citation disadvantage of female scholars indicate that gender-homophily in interpersonal networks (Jadidi, Karimi, Lietz, & Wagner, 2018; Karimi, Génois, Wagner, Singer, & Strohmaier, 2018), gender-homophily in men's citation practices (Dworkin et al., 2020), male-dominated online science dissemination (Procter et al., 2010; Sugimoto, Work, Larivière, & Haustein, 2017; Vásárhelyi, Zakhlebin, Milojević, & Horvát, 2021), and the asymmetrical nature of collaborations between men and women (Langrock & González-Bailón, 2020) could explain the observed citation penalty.

Female-female teams in Engineering benefit more from online visibility compared to male-male teams. The rarity of female-female teams in this research area might contribute to their increased online visibility, which could help with accumulating citations. In Social Sciences, where women are represented best (making up 34% of scholars who published in 2012), we find no interaction between team gender composition and visibility, indicating that higher online visibility benefits everyone, regardless of the team setting. Finally, we also find that teams with mixed gender first/last author configurations (FM and MF) are associated with more citations. This result adds to the ongoing debate about the positive vs negative impact of diversity on team performance (Aggarwal & Woolley, 2013; G. Campbell, Mehtani, Dozier, & Rinehart, 2013; Joshi & Roh, 2009; Kang, Yang, & Rowley, 2006; Kanter, 1977; Rock & Grant, 2016; Tower, Plummer, Ridgewell, & others, 2007). Specifically, Social Sciences are a relatively more gender-balanced research area. Our finding is aligned with previous results that in gender-balanced environments, diversity is more beneficial (Bear & Woolley, 2011; Horwitz & Horwitz, 2007; Joshi, 2014).

Considering the main finding about the link between online visibility at the time of publication and citation impact years later, it is essential to invest in campaigns and tools that can help women (and other underrepresented groups in academia) reach a wider audience online as this could lead to more equitable citations. Some feminist efforts, such as 500 Women Scientists and Art+Feminism (Langrock & González-Bailón, 2020), have intervened successfully and increased women's visibility. Therefore, we encourage institutions and the informetric and scientometric research communities to continue to promote the online visibility of female and other underrepresented scientists' work.